\documentclass[a4paper]{article}

\usepackage{INTERSPEECH2022}
\usepackage{hyperref}
\usepackage{multirow}
\usepackage[marginal]{footmisc}
\usepackage{tabularx}
\usepackage{graphicx}
\usepackage{bbding}
\usepackage{float}
\usepackage{tablefootnote}
\usepackage{siunitx}
\usepackage{threeparttable}

\renewcommand{\footnote}{}

\DeclareMathOperator{\MHA}{MHA}
\DeclareMathOperator{\SMHA}{SMHA}
\DeclareMathOperator{\SMHAS}{SMHAS}
\DeclareMathOperator{\STATS}{STATS}

\title{The Kriston~AI System for the VoxCeleb Speaker Recognition Challenge 2022}
\name{Qutang Cai$^1$\textsuperscript{\dag}, Guoqiang Hong$^1$\textsuperscript{\ddag}, Zhijian Ye$^1$, Ximin Li$^1$, Haizhou Li$^{1,2,3}$}
\address{
  $^1$Kriston AI Lab, China\\
  $^2$The Chinese University of Hong Kong, Shenzhen, China \\
  $^3$Department of Electrical and Computer Engineering, National University of Singapore, Singapore
}
\email{\{caiqt, hgq, yzj, lixm, haizhou.li\}@kuaishang.com.cn}
\begin{document}

\maketitle

\begingroup
\renewcommand\thefootnote{\dag}
\footnotetext{Corresponding author.}
\renewcommand\thefootnote{\ddag}
\footnotetext{Main contributor for track~4.}
\endgroup
\renewcommand\thefootnote{1}

\begin{abstract}
  This technical report describes our system for track~1,~2 and~4 of the VoxCeleb Speaker Recognition Challenge 2022~(VoxSRC-22).
  By combining several ResNet variants, our submission for track~1 attained a $\mathrm{minDCF}$ of $0.090$ with EER $1.401\%$. By further incorporating three fine-tuned pre-trained models, our submission for track~2 achieved a $\mathrm{minDCF}$ of $0.072$ with EER $1.119\%$. For track~4, our system consisted of voice activity detection (VAD), speaker embedding extraction, agglomerative hierarchical clustering (AHC) followed by a re-clustering step based on a Bayesian hidden Markov model and overlapped speech detection and handling. Our submission for track~4 achieved a diarisation error rate (DER) of 4.86\%. The submissions all ranked the 2nd places for the corresponding tracks.

\end{abstract}

\section{Introduction}
The VoxSRC-22 challenge contains two full supervised speaker verification tracks (track~1 and track~2), and one diarisation track (track~4), where
\begin{description}
  \item[track~1] is a closed task, and only VoxCeleb2~\cite{vox2} dev dataset can be used for training models;
  \item[track~2 and ~4] are both open tasks, and any public data except the challenge test data can be used.
\end{description}
The goal of this challenge is to probe how well current methods can segment and recognize speakers from speech obtained 'in the wild'.

For track~1, we trained from scratch six models modified from the ResNet~\cite{resnet} architecture, using only VoxCeleb2~\cite{vox2} dev dataset.
For track~2, we additionally fine-tuned three recently proposed pre-trained models~\cite{xlsr_2021,wavlm_2021}, which are all publicly available, to harness the power of the large-scale pre-trained models.
All the models in track~1 and~2 were trained and calibrated individually with the same procedure, and then fused using weighted linear combinations.

\begin{figure}[ht]
  \centering
  \includegraphics[width=\linewidth]{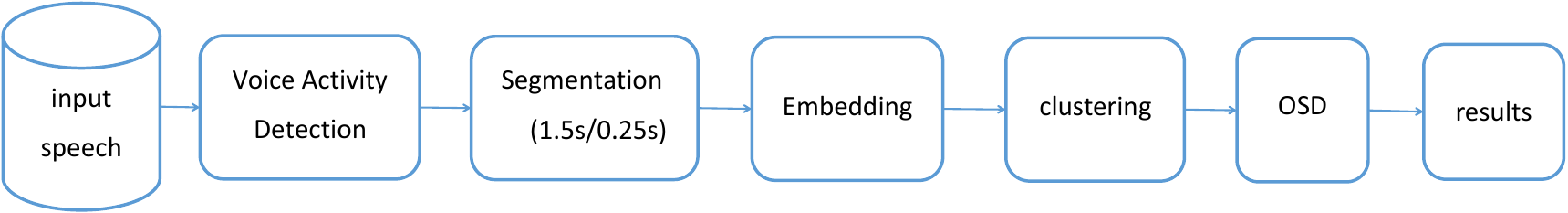}
  \caption{Diarisation system overview.}
  \label{fig:diar_overview}
\end{figure}

For track~4, we built our speaker diarization system by means of VAD, speaker embedding extraction, clustering, overlapped speech detection (OSD) and handling, step by step as shown in Figure~\ref{fig:diar_overview}.

\section{Data preparation and augmentation}
\subsection{Training data}
\textbf{Track~1\&2:}
For training, we used the VoxCeleb2~dev dataset  which contains 1,092,009 utterances and
5,994 speakers in total.
Moreover, we employed a speaker augmentation strategy with 3-fold speed augmentation~\cite{speaker_aug_2019,speaker_aug_2020,speakin_vox21} and thus obtained 17,982 speakers.
Besides, data augmentation for training was carried out in an online manner, with the Kaldi-style augmentation~\cite{Povey_ASRU2011}, including MUSAN~\cite{musan} noises, music, and babble and reverberation from the Room Impulse Response and Noise Database (RIR)\cite{rirs}.

For validation, four development sets were used, including VoxCeleb1-O, VoxCeleb1-E, VoxCeleb1-H~\cite{VoxCeleb1_interspeech}  and VoxSRC22-dev\textsuperscript{1}\footnotetext{We used the cleaned trial lists of VoxCeleb1-O, -E and -H.}.

\noindent \textbf{Track~4:}
The developement set consisted of the development set and test set of VoxConverse\cite{2020Spot}(for convenience, in the following parts, they are referred to as track4-dev1 and track4-dev2, respectively, and the evaluation set is referred to as track4-test). The datasets used in this challenge for each model are described as follows:
\begin{itemize}
\item VAD: NIST(LDC2009E100)\cite{21sre}, LibriSpeech\cite{2015Librispeech}, AISHELL-2\cite{2018AISHELL}, the noise of track4-dev1 and track4-dev2 were the mixed
training set. We used track4-dev2 for validation.
\item AHC: We directly tuned the parameters on track4-dev2.
\item Variational Bayes hidden Markov model clustering: We directly tuned the parameters on track4-dev2.
\item OSD:  NIST(LDC2009E100), LibriSpeech, AISHELL-2 were used as the mixed
training set. We used track4-dev2 for validation.
\item Data augmentation: We performed data augmentation with MUSAN and RIRs corpus.
\end{itemize}

\subsection{Features}
\noindent \textbf{Track~1\&2:}
For track~1, we used mean normalized Kaldi-compliant log Mel-filter bank (FBank) features with energies with a 25~ms window size and a 10~ms frameshift. The feature dimensions were chosen from $\{96, 104, 112, 120\}$ in our experiments.
For fine-tuning models in track~2, we directly used the raw waveform. No additional voice activity detection (VAD) was used throughout this report.

\noindent \textbf{Track~4:}
For VAD and OSD, we used mean normalized Kaldi-compliant 80-dim FBank and 30-dim MFCC features with energies with a 25~ms window size and a 10~ms frameshift.

\section{System description for track~1 and~2}
\subsection{Model architectures: track~1}\label{sec:model:track1}
\begin{figure}[h]
	\begin{center}		
    \includegraphics[width=0.85\linewidth]{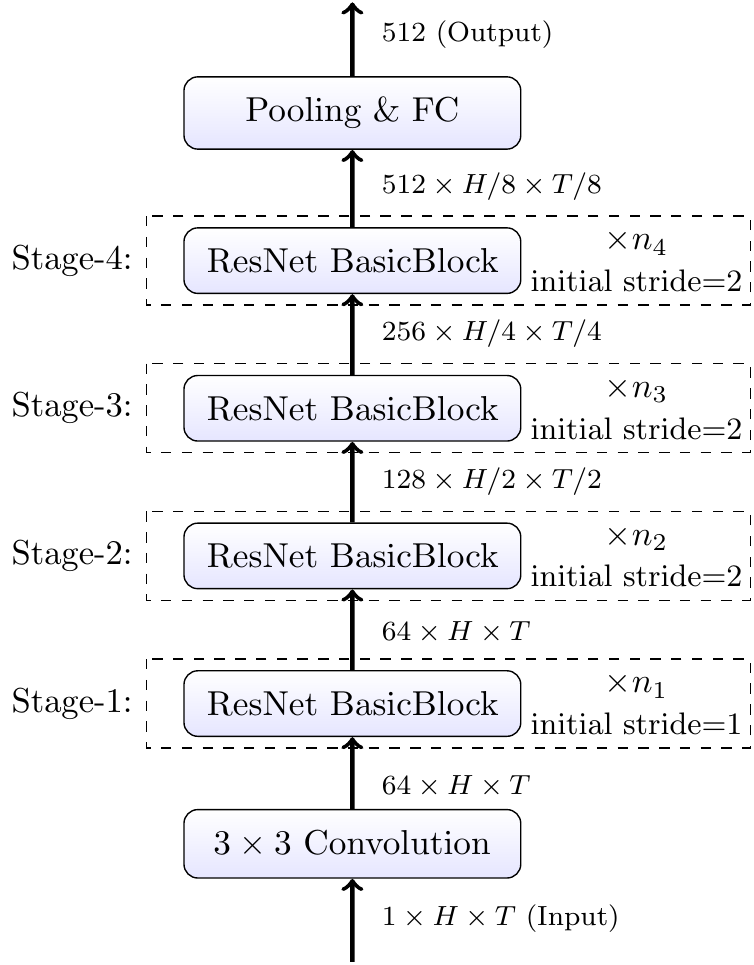}
	\end{center}
	\caption{Base ResNet architecture.}
	\label{fig:resnet}
\end{figure}

\noindent \textbf{ResNet variants}:\quad
The models for track~1 were based on the ResNet architecture which is depicted in Figure~\ref{fig:resnet}, whose base channels were fixed to 64. Moreover, we only considered the basic Resnet block used in ResNet34~\cite{resnet}. 
We modified the ResNet architecture with one or more of the strategies listed in Table~\ref{tbl:modifying_strategy} to introduce modelling diversity, and the resulting models are listed in Table~\ref{tbl:resnet_variants}. In Table~\ref{tbl:modifying_strategy}:
\begin{itemize}
  \item We only applied \textbf{M3} and \textbf{M4} to the first two stages of the backbone due to memory limits and the suggestions in ~\cite{asru21_se}.
  \item For \textbf{M4}, we used channel-wise and frequency-wise squeeze-excitation in ~\cite{ecapa_cnn_tdnn,asru21_se} to the residual connection, simultaneously. It's worth mentioning that we additionally introduced bias items to the input which also depend on the input like the weights items. 
      
  \item For \textbf{M5}, we altered the downsampling operation at the beginning of each stage from a 2-stride $2\times 2$ convolution with a $2\times 2$ average pooling operation.
\end{itemize}

\begin{table}[th]
  \centering
  \begin{tabularx}{0.9\linewidth}{ l l}
    \toprule
    {\textbf{Name}} & {\textbf{Description}} \\
    \midrule
    \textbf{M1} & Changing input feature dimension \\
    \textbf{M2} & Changing model depths \\
    \textbf{M3} & Changing kernel sizes \\
    \textbf{M4} & Using attention mechanisms~\cite{ecapa_cnn_tdnn,asru21_se} \\
    \textbf{M5} & Using other downsampling operations~\cite{tricks_cvpr19} \\
    \bottomrule
  \end{tabularx}
  \caption{Strategies for modifying ResNet.}
  \label{tbl:modifying_strategy}
\end{table}

\begin{table}[th]
  \centering
  \begin{tabularx}{0.9\linewidth}{ l c c l l l}
    \toprule
    {\textbf{Name}} & {\textbf{M1}} & {\textbf{M2}} & {\textbf{M3}} & {\textbf{M4}}  & {\textbf{M5}}\\
    \midrule
    \textbf{R1} & 96 & $3\times 6 \times 20 \times 3$ & \XSolidBrush  &  \Checkmark & \XSolidBrush\\ 

    \textbf{R2} & 112 & $3\times 5 \times 14 \times 3$ & \XSolidBrush  &  \Checkmark & \XSolidBrush \\ 

    \textbf{R3} & 120 & $3\times 6 \times 14 \times 3$ & \XSolidBrush &  \Checkmark & \XSolidBrush \\ 

    \textbf{R4} & 104 & $3\times 5 \times 16 \times 3$ & \XSolidBrush  &  \Checkmark & \Checkmark \\ 

    \textbf{R5} & 104 & $3\times 4 \times 16 \times 3$ & 9  &  \Checkmark &  \Checkmark\\ 

    \textbf{R6} & 96 & $3\times 5 \times 16 \times 3$ & 9  &  \Checkmark &  \Checkmark \\ 

    \bottomrule
  \end{tabularx}
  \caption{ResNet variants for Track~1.}
  \label{tbl:resnet_variants}
\end{table}

\noindent \textbf{Pooling layer}:\quad
Based on the multi-head attention ($\MHA$) pooling ~\cite{mhasp_interspeech}, we propose a \textbf{s}huffled \textbf{m}ulti-\textbf{h}ead \textbf{a}ttention ($\SMHA$) pooling method.
Note that in $\MHA$, for each channel groups of the input, the forward step works independently without any interaction between each other.
However, we believed it could be better to introduce interaction between the heads, so we applied the shuffle operation in ~\cite{shufflenet} and carried out the pooling operation as:
\begin{equation}\label{eqn:SMHAP}
  \SMHA (x) = \MHA \Big( \mathrm{CAT} \big(x, \mathrm{SHUFFLE}(x) \big) \Big)
\end{equation}
where $\mathrm{CAT}$ is the concatenation operation.
Additionally, when calculating the attention weights using $\MHA$ in $\SMHA$, we observed improvements if each head's statistics vector (its mean and standard deviation) were also considered. We name this variant of $\SMHA$ as \textbf{s}huffled \textbf{m}ulti-\textbf{h}ead \textbf{a}ttention with \textbf{s}tatistics ($\SMHAS$), which is used for the ResNet Variants throughout this report. In our experiments, all the head numbers were fixed to 8.

\subsection{Model architectures: track~2}
The models for Track~2 consisted of the models for Track~1 (see also \nameref{sec:model:track1}) and three fine-tuned pre-trained models, including WavLM~Large (WavLM-L)~\cite{wavlm_2021}, Facebook's Wav2Vec2 XLS-R 300M (XLSR-300M) and 1B (XLSR-1B)~\cite{xlsr_2021}. The hidden states of the pre-trained models were extracted using S3PRL\textsuperscript{1}\footnotetext{\url{https://github.com/s3prl/s3prl}}, and then normalized, linear weight combined, and fed to a downstream model similar to~\cite{wavlm_2021}, where the downstream model was ECAPA-TDNN~\cite{ecapa_2000} with 1024 base channels and a 512-dimensional output. The resulting models are listed in Table~\ref{tbl:pretrained_models}, where $\STATS$ means the statistics pooling layer~\cite{xvector}.

\begin{table}[th]
  \centering
  \begin{tabularx}{0.75\linewidth}{ l l c}
    \toprule
    {\textbf{Name}} & {\textbf{Upstream model}} & {\textbf{Pooling layer}} \\
    \midrule
    \textbf{P1} & WavLM-L & $\SMHA$ \\ 

    \textbf{P2} & XLSR-300M &  $\STATS$ \\ 

    \textbf{P3} & XLSR-1B & $\STATS$ \\ 

    \bottomrule
  \end{tabularx}
  \caption{Fine-tuned pretrained models.}
  \label{tbl:pretrained_models}
\end{table}

\begin{table*}[h]
  \centering
  \caption{Single system evaluation results.}\label{tbl:eval:vox22}
  \begin{threeparttable}
  \setlength{\tabcolsep}{1.5mm}{
  \begin{tabular}{lcccccccc}
    \toprule
    \multirow{4}{*}{\textbf{System}} &
    \multicolumn{2}{c}{\multirow{2}{*}{\textbf{VoxCeleb1-O}}} &
    \multicolumn{2}{c}{\multirow{2}{*}{\textbf{VoxCeleb1-E}}} &
    \multicolumn{2}{c}{\multirow{2}{*}{\textbf{VoxCeleb1-H}}} &
    \multicolumn{2}{c}{\multirow{2}{*}{\textbf{VoxSRC22-dev}}}
    \\ \\
    \cline{2-9}
    & \multirow{2}{*}[-2pt]{\textbf{EER(\%)}} &  \multirow{2}{*}[-2pt]{$\textbf{DCF}_\textbf{0.05}$} & \multirow{2}{*}[-2pt]{\textbf{EER(\%)}} & \multirow{2}{*}[-2pt]{$\textbf{DCF}_\textbf{0.05}$} &
    \multirow{2}{*}[-2pt]{\textbf{EER(\%)}} & \multirow{2}{*}[-2pt]{$\textbf{DCF}_\textbf{0.05}$} &
    \multirow{2}{*}[-2pt]{\textbf{EER(\%)}} & \multirow{2}{*}[-2pt]{$\textbf{DCF}_\textbf{0.05}$} \\ \\
    \midrule
\textbf{R1} & 0.3510 & 0.0220 & 0.6077 & 0.0321 & 0.9866 & 0.0545 & 1.5691 & 0.1110\\
\textbf{R2} & 0.3776 & 0.0244 & 0.5860 & 0.0318 & 0.9131 & 0.0521 & 1.5350 & 0.1109\\
\textbf{R3} & 0.3616 & 0.0241 & 0.6205 & 0.0333 & 0.9687 & 0.0560 & 1.5556 & 0.1123\\
\textbf{R4} & 0.3457 & 0.0299 & 0.5739 & 0.0312 & 0.9031 & 0.0511 & 1.5186 & 0.1070\\
\textbf{R5} & 0.3829 & 0.0271 & 0.5788 & 0.0321 & 0.8944 & 0.0499 & 1.5002 & 0.1071\\
\textbf{R6} & 0.3297 & 0.0272 & 0.5771 & 0.0315 & 0.9012 & 0.0512 & 1.5099 & 0.1072\\
\textbf{P1} & 0.3615 & 0.0327 & 0.4705 & 0.0278 & 0.9578 & 0.0582 & 1.4591 & 0.1000\\
\textbf{P2} & 0.5797 & 0.0523 & 0.4977 & 0.0296 & 0.9045 & 0.0539 & 1.4140 & 0.0899\\
\textbf{P3} & 0.5159 & 0.0434 & 0.4525 & 0.0286 & 0.8759 & 0.0542 & 1.4163 & 0.0962
    \\ \midrule
Fusion &  &  &  &  &  &  &  & \\
    \midrule
\textbf{track1}& 0.2393 &  0.0209 & 0.4974 & 0.0266 & 0.8160 & 0.0452 & 1.3598 & 0.0977 \\
\textbf{track2} & 0.2021 & 0.0153 & 0.3481 & 0.0286 & 0.6262 & 0.0354 & 1.0468 & 0.0760 \\
    \bottomrule
  \end{tabular}}
  \end{threeparttable}
\end{table*}

\subsection{Training procedure}
A two-stage training procedure like ~\cite{speakin_vox21,icassp_voxsrc20} was adopted for training the models:
\begin{description}
  \item[Stage-1] Train initial models using short utterances to speedup the training process, where the short utterances were randomly cropped from the corresponding original ones with 2 and 2.24 seconds, respectively for track~1 and track~2.
      The loss function used in this stage was AM-Softmax with subcenters and inter-topK penalties (SC-ITK-AMSoftmax) ~\cite{speakin_vox21,amsoftmax}, with subcenter number=3, margin=0.2, scale=35, inter-topK neighbor size=5, and inter-topK penalty=0.06.

  \item[Stage-2] Train final models with the large margin fine-tuning (LMF~\cite{icassp_voxsrc20}) technique, removing the speaker augmentation and using longer utterances with 6 seconds to match the target domain, while for short speech segments wrap padding were used. The loss function used here was AAM-Softmax with subcenters (SC-AAMSoftmax)~\cite{arcface_cvpr2019}, with subcenter number=3, margin=0.5, scale=35.
\end{description}

Throughout the training processes, each epoch contained 3,000 iterations, and the batch sizes were set to 384 and 128 when possible\textsuperscript{1}\footnotetext{Gradient accumulation technique was used to catch up when we were confronted with the hardware memory limits.}, respectively, for Stage-1 and~2. We used AdamW (with weight decay 0.0001) as the optimizer, and a ReduceLROnPlateau scheduler as the learning rate scheduler (with updating
frequency 3,000, patience 4, and decaying factor 0.4).  For the ResNet variants, the start learning rates were \num{3e-4}
 and \num{4e-5} for Stage~1 and~2, respectively. For fune-tuning the pre-trained models, the situation was slightly more complicated and required special treatment due to the huge model sizes, and the details is described in the following section.

\subsection{Fine-tuning pre-trained models}
The basic fine-tuning steps are carried out as follows:
\begin{itemize}
  \item For \textbf{P1} and \textbf{P2}, we took the following three steps for model training in Stage-1:
  \begin{description}
    \item[Step-1] Freezing the upstream models, train the downstream models, with a start learning rate of \num{3e-4}.
    \item[Step-2] Unfreezing the upstream models and freezing the downstream models, train the upstream models, with a start learning rate of \num{4e-5}.
    \item[Step-3] Unfreezing the whole model parameters, train the entire models, with a start learning rate of \num{4e-5}.
  \end{description}
    In Stage-2, we trained the entire models with a start learning rate of \num{2e-5}.
  \item For \textbf{P3}, we were hindered by the hardware memory limits; consequently, we trained only its self attention weights and the downstream model, alternatively. The training steps in Stage-1 are described as follows:
      \begin{description}
        \item[Step-1] Freezing the upstream model, train the downstream model, with a start learning rate of \num{3e-4}.
        \item[Step-2] Train the self attention weights (in the upstream model) and the downstream model alternatively for two cycles:
        \begin{description}
          \item[Step-2.1] Freezing the model parameters except the self attention parts, train the self attention weights with a start learning rate of \num{4e-5}.
          \item[Step-2.2] Freezing the upstream model, train the downstream model with a start learning rate of \num{3e-4}.
        \end{description}
      \end{description}
      The training steps in Stage-2 were also carried out similarly, training the self attention weights and the downstream model alternatively, except that the start learning rates were all set to \num{2e-5}.
\end{itemize}
However, we had observed the tendency of overfit when fine-tuning the pre-trained models. Therefore, we saved model checkpoints after each epoch finished, and picked the one that performed best on the validation set for the final system.

\subsection{Scoring procedure}
When extracting the speaker embedding vectors, the $L_2$-normalized 512-dimensional outputs of the last full connected layer of each model were used.
When performing single system scoring, we computed the cosine similarity score of the speaker embeddings of each trial, and then used adaptive score normalization (AS-Norm)~\cite{s_norm,s_norm_2} and quality measure functions (QMF)~\cite{icassp_voxsrc20,quality_cal_2013} for calibration.
For building cohorts used in AS-NORM, we randomly picked at most 30 utterances for each speaker from the VoxCeleb2 dev dataset without augmentation, extracted their embeddings, and then averaged them speaker-wisely; the resulting vectors were used as the cohorts, in which only top 300 imposter scores were used for score normalization. The calibration  was trained on the VoxCeleb1-H trials using logistic regression in a similar way to ~\cite{speakin_vox21,icassp_voxsrc20}.

The final system was a linear weighted combination of the individual calibrated models. The combination weights were picked manually: for both tracks, each weight for \textbf{R1}---\textbf{R6} was simply set to 1; for track~2, the weights of \textbf{P1}, \textbf{P2} and \textbf{P3} were set to 1, 1 and 2, respectively.

\section{System description for track~4}
\subsection{Overview}
The proposed speaker diarisation system is illustrated in Figure~\ref{fig:diar_overview}. The input audio was first processed by VAD to obtain valid speech segments. Then speaker embeddings were extracted with a 1.5s sliding window size with 0.25s step size. Clustering and OSD were conducted individually. The details are explained in the following subsections.

\subsection{Voice activity detection}
We trained two VAD models like \cite{2021The} except that we used different acoustic features, including 30-dim MFCC and 80-dim FBank.
In addition, the VAD functionality provided by pyannote 2.0\cite{2020Pyannote} was also included as a sub-system. We then adopted a multi-system fusion method as \cite{2020Microsoft}, and combined these three sub-systems with equal weights. Table~\ref{tab:vad} shows the false alarm (FA) and miss detection (MISS) on track4-dev2.

\begin{table}[ht]
  \caption{The false alarm (FA), miss detection (MISS) and accuracy of the VAD model.}
  \label{tab:vad}
  \centering
  \begin{tabular}{lccc}
    \toprule
    \textbf{System}      & \textbf{FA[\%]}   & \textbf{MISS[\%]}     & \textbf{Accuracy[\%]}          \\
    \midrule
    FBank                & 3.49              & 1.49                  & 95.00                         \\
    MFCC                 & 4.27              & 0.92                  & 94.80                          \\
    pyannote             & 3.22              & 1.62                  & 95.15                         \\
    Fusion               & 3.55              & 1.06                  & 95.37                         \\
    \bottomrule
  \end{tabular}
\end{table}

\subsection{Speaker Embedding}
We used model R6 in Table~\ref{tbl:resnet_variants} which achieved an EER=0.44\% using cosine similarity on VoxCeleb1-O.

\subsection{Clustering}
We performed AHC on audio segments, and then performed an re-clustering step based on the Bayesian hidden Markov model.

\subsubsection{Initial Clustering}
The speaker embeddings were clustered by means of AHC\cite{2020Analysis} with cosine similarity. The AHC clustering threshold was tuned on track4-dev2, combined with Variational Bayes hidden Markov model (VB-HMM) diarisation\cite{2020Bayesian}.

\subsubsection{Re-clustering}
We replaced equation (17) and (18) in VB-HMM\cite{2020Bayesian} by \eqref{eq2} and \eqref{eq3}:

\begin{equation}
  \bm{L_s} = \bm{I} + \frac{F_A}{F_B}\sum_t \gamma_{ts}
  \label{eq2}
\end{equation}

\begin{equation}
  \bm{\rho_t} = \bm{x_t} = F_C\bm{E_t}
  \label{eq3}
\end{equation}
where \(\gamma_{ts}\) is the marginal approximate posterior at frame t for speaker s; \(F_A=0.3\), \(F_B=17\); \(F_C\) is a scale parameter; \(\bm{E_t}\) is the L2-normalized speaker embedding at frame t; \(\bm{I}\) is a vector of 1s.

 We also considered using AS-Norm for score calibration. For building cohorts used in AS-Norm, we randomly picked 2 utterances for each speaker from the VoxCeleb2 dev dataset, cropped them to 1.5 seconds and extracted their embeddings. We then replaced the \(\bm{\alpha_s^{T}}\bm{\rho_t}\) and \(\bm{\Phi}\) terms in equation (23) in \cite{2020Bayesian} by

\begin{equation}
    \bm{\alpha_s^{T}}\bm{\rho_t} = \frac{F_AF_C^{2}}{F_B} l_s^{-1}  \frac{\bm{\beta_s^{T}}\bm{E_t}-\mu_s}{\sigma_s}\sum_t{\gamma_{ts}}
    \label{eq4}
\end{equation}

\begin{equation}
    \bm{\Phi} = \bm{I}
    \label{eq5}
\end{equation}
where \(\bm{\beta_s} = \frac{\sum_t{\gamma_{ts}\bm{E_t}}}{\sum_t{\gamma_{ts}}}\), \(l_s = 1.0 + \frac{F_A}{F_B}\sum_t \gamma_{ts}\), and \(\mu_s\) and \(\sigma_s\) are mean and standard deviation of \(\bm{\beta_s}\).

\begin{table}[ht]
  \caption{The DER and JER of the proposed speaker diarization system on track4-dev2.}
  \label{tab:diar system on dev2}
  \centering
  \begin{tabular}{lcc}
    \toprule
    \textbf{System}    & \textbf{DER[\%]}   & \textbf{JER[\%]}         \\
    \midrule
    VB                 & 4.42              & 26.43                     \\
    VB+asnorm          & 4.29              & 26.81                     \\
    \bottomrule
  \end{tabular}
\end{table}

\subsection{Overlapped speech detection and handling}
The overlap detection model, including its training process, were similar to that of the VAD model. We trained two models with the same structure and fused with pyannote 2.0. For each overlapped speech segments, we found the two closest speakers in time.

\section{Experimental results}
\subsection{Track~1\&2}
We provide in Table~\ref{tbl:eval:vox22} the single system results evaluated on the validation trial lists. The results in Table~\ref{tbl:eval:vox22} show that although the single system performances are close to each other, the fused system's can still achieve a considerable improvement, which also indicates the effectiveness of utilizing  the diversities of the single systems.
On the test trials of this challenge, the fused system achieved a minDCF of 0.090 and an EER of 1.401\% for track~1, and achieved a minDCF of 0.072 and an EER of 1.119\% for track~2, where the testing results were all closed to the validation results on the VoxSRC22-dev dataset.

\subsection{Track~4}
The diarisation results of the proposed systems are shown in Table~\ref{tab:diar system on dev2}. The system VB+asnorm was our best system. Compared with the system VB, DER was improved by 0.13\%, but the JER was deteriorated by 0.38\%. Our best submission on the evaluation set attained DER 4.86\% and JER 25.48\%.

\bibliographystyle{IEEEtran}
\bibliography{voxsrc22}
\end{document}